\documentclass[twocolumn,showpacs,preprintnumbers,amsmath,amssymb]{revtex4}

\usepackage{graphicx}
\usepackage{dcolumn}
\usepackage{bm}

\begin{document}
\def\brho{{\hbox{\boldmath $\rho$}}}
\def\bsb{{\hbox{\boldmath $\beta$}}}
\def\bsk{{\hbox{\boldmath $k$}}}
\def\bsp{{\hbox{\boldmath $p$}}}

\title{Features of two-pion Bose-Einstein correlations based on
event-by-event analysis in smoothed particle hydrodynamics}

\author{Yan-Yu Ren$^1$}
\author{Wei-Ning Zhang$^{1,2}$\footnote{wnzhang@dlut.edu.cn}}
\author{Jian-Li Liu$^{1}$}

\affiliation{
$^1$Department of Physics, Harbin Institute of Technology,
Harbin, Heilongjiang 150006, China\\
$^2$School of Physics and Optoelectronic Technology, Dalian University
of Technology, Dalian, Liaoning 116024, China\\
}

\date{\today}

\begin{abstract}
We examine the space-time structure of the particle-emitting sources
with fluctuating initial conditions in smoothed particle
hydrodynamics.  The two-pion correlation functions of single events
for the sources exhibit event-by-event fluctuations.  The large
event-by-event fluctuations and wide distributions of the
error-inverse-weighted fluctuations between the HBT correlation
functions of single and mixed events are important features for the
sources with the fluctuating initial conditions.  The
root-mean-square of the weighted fluctuations is a signal to detect
the inhomogeneity of the systems produced in high energy heavy ion
collisions.
\end{abstract}


\maketitle

\section{Introduction}

The main physics goal of heavy ion collisions at the BNL Relativistic
Heavy Ion Collider (RHIC) is the study of the extremely hot and
dense matter, the quark-gluon plasma (QGP), formed in the early
stage of the collisions. Hydrodynamics may provide a direct link
between the early state of the matter and final observables and has
been extensively used in high energy heavy ion collisions.  The
hydrodynamic calculations with the equation of state of the QGP
agree well with the RHIC $v_2$ data of the elliptic flow at low
transverse momentum $p_{_T}<2$ GeV
\cite{STA01,STA02,STA03,PHE03,STA04}, which is believed as an
important evidence of the existence of a strongly coupled QGP in the
early stages of the collisions \cite{Gyu04,BRA05,PHO05,STA05,PHE05}.
However, hydrodynamic results can not explain the RHIC
Hanbury-Brown-Twiss (HBT) measurements, $R_{\rm out}/R_{\rm side}
\approx 1$ \cite{STA01a,PHE02a,PHE04a,STA05a}.  It is the so-called
HBT puzzle.

In Ref. \cite{Zha04} a granular source model of QGP droplets
evolving hydrodynamically was put forth to explain the HBT puzzle.
The suggestion was based on the observation that in the hydrodynamic
calculations for the granular source the average particle emission
time scales with the initial radius of the droplet, whereas the
spacial size of the source is the scale of the distribution of the
droplets.  For a granular source with many of the small droplets
distributed in a relatively large region, the HBT radius $R_{\rm
out}$ can be close to $R_{\rm side}$ \cite{Zha04}. In Ref.
\cite{Zha06} the authors further investigated the elliptic flow and
HBT radii as a function of the particle transverse momentum for an
improved granular source model of QGP droplets.  They argued that
although a granular structure was suggested earlier as the signature
of a first-order phase transition
\cite{Wit84,Pra92,Cse92,Zha95,Ala99,Ran04,Zha04,Won04,Zha05}, the
occurrence of granular structure may not be limited to first-order
phase transition \cite{Zha06,Zha07}. In an event-by-event basis, the
initial transverse distribution of energy density in nucleus-nucleus
collisions have been known to be highly fluctuating
\cite{Gyu97,Dre02,Ham04}.  This large spatial fluctuation may
facilitate the occurrence of instability of the system during it
violent expansion subsequently and the fragmentation to many
granular droplets together with surface tension effect
\cite{Zha06,Zha07}. So, the examination of the space-time structure
in event-by-event basis is important to understand the system
evolution and the HBT puzzle.

Smoothed Particle Hydrodynamics (SPH) is a suitable candidate that
can be used to treat the system evolution with large fluctuating
initial conditions for investigating event-by-event attributes in
high energy heavy ion collisions \cite{Agu01,Ham04}.  The main idea
of SPH is the parametrization of the fluid in terms of discrete
Lagrangian coordinates attached to small volumes (call `particles')
with some conserved quantities in hydrodynamics
\cite{Agu01,Ham04,Han05}.  NEXSPHERIO is a SPH code
\cite{Gaz03,Ham04,And06} with event-by-event initial conditions
generated by NEXUS event simulator \cite{Dre02}.  It has been used
to study a wide range of problems in high energy heavy ion
collisions \cite{Agu01,Gaz03,Ham04,Soc04,Ham05,And06,Qia07}. In this
article we use NEXSPHERIO to simulate the evolution of the system
produced in the collisions of $\sqrt{s_{NN}}=200$ GeV Au+Au at RHIC.
We will examine the space-time structure of the systems and
investigate the two-pion HBT correlation functions in event-by-event
basis. Our results show that the systems are inhomogeneous in space
and time both for a first-order and a cross-over transitions between
the QGP and hadronic phase.  There are ``lumps" in different regions
of the system.  The two-pion HBT correlation functions in
event-by-event basis exhibit large fluctuations, which lead to a
wide distribution of the error-inverse-weighted fluctuations between
the HBT correlation functions of single and mixed events.  These large
fluctuations and wide distributions are important features for the
sources with event-by-event fluctuating initial conditions.  The
root-mean-square of the weighted fluctuations is a signal to
detect the inhomogeneity of the systems produced in high energy
heavy ion collisions.

\section{System evolution and space-time structure}

In hydrodynamics the behavior of system evolution is determined by
the initial conditions and the equation of state (EOS) of the
system.  The system initial states of NEXSPHERIO are given by the
NEXUS code, which can provide detailed space distributions of
energy-momentum tensor, baryon-number, strangeness, and charge
densities at a given initial time (which is taken to be $\tau_0=1$
fm in our calculations), in event-by-event basis \cite{Dre02,Ham04}.
In our calculations we use two kinds of EOS. The first one, EOS-I,
considers a first-order phase transition at $T_c=160$ MeV between a
QGP and a hadronic resonance gas as used in Refs.
\cite{Gaz03,Soc04}.  The QGP is an ideal gas of massless quarks (u,
d, s) and gluons and the hadronic gas contains the resonances with
mass below 2.5 GeV/c$^2$, where volume correction is taken into
account \cite{Gaz03,Soc04}. Because at RHIC energy the net baryon
constituent is very small and the phase transition is a smooth cross
over, in the mid-rapidity region of the heavy ion collisions, a
modification for the EOS with first-order phase transition in the
hydrodynamical code should be considered \cite{Ham04,Ham05}.
Accordingly, we introduce EOS-II which is obtained by smoothing the
EOS-I in the transition region with the entropy density suggested by
QCD lattice results \cite{Bla87,Lae96,Ris96,Zha04,Yul08}.  Fig. 1
shows the relations of the thermodynamical quantities for the systems
evolving with the EOS-I (grey) and the EOS-II (black), where
$s,\,\epsilon$, and $P$ are entropy density, energy density, and
pressure of the system. The width of the transition $\Delta T$ for
the EOS-II is taken as $0.1 T_c$ \cite{Ris96}.

\begin{figure}
\includegraphics[angle=0,scale=0.39]{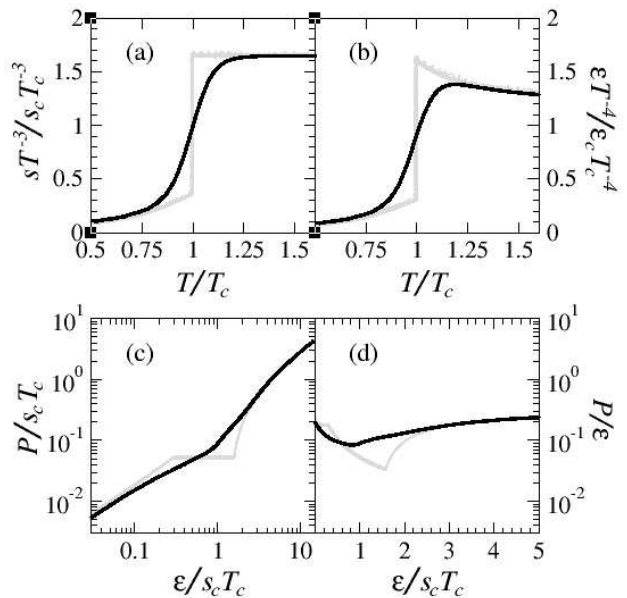}
\caption{\label{fig:fig1} (a) entropy density $s$, (b) energy density
$\epsilon$, (c) pressure $P$, and (d) $P/\epsilon$ of the systems
evolving with EOS-I (grey) and EOS-II (black). }
\end{figure}

The coordinates used in NEXSPHERIO are $\tau=\sqrt{t^2-z^2},\,x,\,y,$
and $\eta=(1/2)\ln [(t+z)/(t-z)]$ \cite{Agu01,Ham04,Gaz03}.  They
are convenient for a system with rapid longitudinal expansion.
However, in order to examine the space-time structure of the system 
with nonlocal coordinates we work in the center-of-mass frame of the
system.  Figs. 2(a), (b), (c), and (d) are the pictures of the transverse 
distributions of energy density for one NEXSPHERIO event ($\sqrt{s_{NN}}
=200$ GeV Au+Au) with impact parameter $b=0$ fm and evolving with EOS-I.  
The pictures are taken for the region ($|x,y|<12$ fm and $|z|<1$ fm) and
at $t=1,\,5,\,9$, and 13 fm$/c$ with an exposure of $\Delta t=0.3$
fm$/c$.  Figs. 2(a$'$), (b$'$), (c$'$), and (d$'$) are the pictures
for the event with the same initial conditions but evolving with
EOS-II.  One can see that the systems are inhomogeneous in space and
time both for the events evolving with the EOS-I and the EOS-II.
There are many ``lumps" in the systems. By comparing the pictures
for different events we find that the lump locations are different
event-by-event.  Figs. 3((a)--(d)) and ((a$'$)--(d$'$)) show the
pictures of the transverse distributions of energy density for the
events with different impact parameters and evolving with EOS-I and
EOS-II, respectively. One can see that the number of the lumps in
the system decreases with impact parameter increasing. Because the
systems evolving with the two kinds of EOSs have the similar
space-time structure, we will consider only the EOS-II in our
calculations later.

\begin{figure}
\includegraphics[angle=0,scale=0.80]{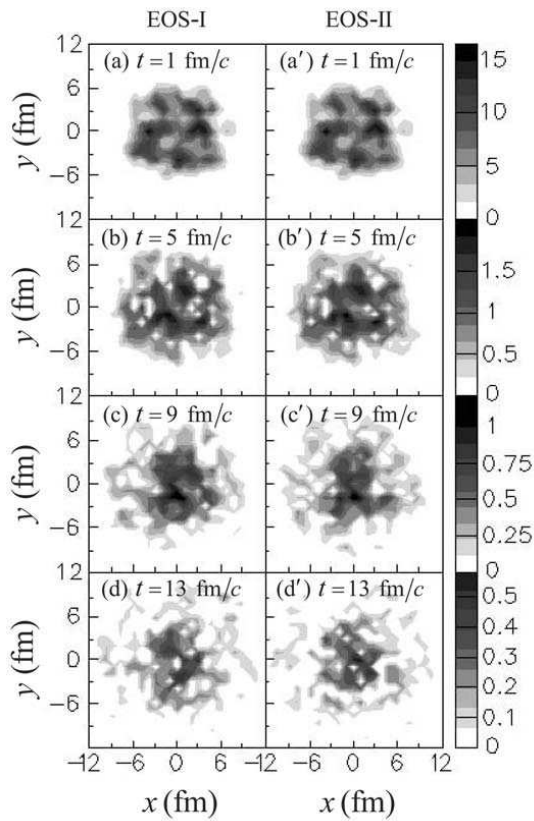}
\caption{\label{fig:fig2} Transverse distributions of energy density of
the events evolving with EOS-I ((a)--(d)) and EOS-II ((a$'$)--(d$'$)) at
different times for $\sqrt{s_{NN}}=200$ GeV Au+Au collisions, $b=0$ fm.
The unit of energy density is GeV/fm$^3$.}
\end{figure}

\begin{figure}
\includegraphics[angle=0,scale=0.78]{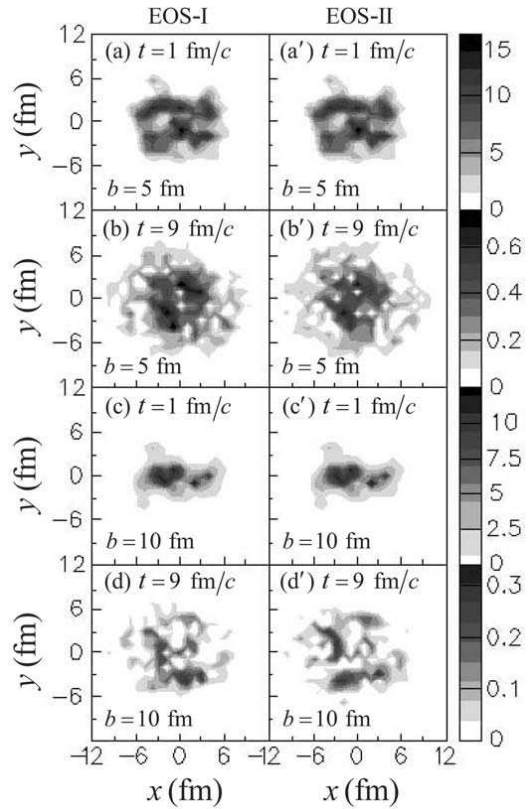}
\caption{\label{fig:fig3} Transverse distributions of energy density of
the events for $\sqrt{s_{NN}}=200$ GeV Au+Au collisions with $b=5$ and
10 fm.  The systems evolve with EOS-I ((a)--(d)) and EOS-II
((a$'$)--(d$'$)).  The unit of energy density is GeV/fm$^3$.}
\end{figure}

\section{Two-pion correlation functions in event-by-event basis}

Two-pion interferometry is a powerful tool for probing the space-time
structure of particle-emitting source.
For the source with dense and void density oscillations, the single-event
HBT correlation functions will appear fluctuations relative to the
mixed-event HBT correlation functions \cite{Won04,Zha05}.  The fluctuation
patterns are different event-by-event.

Assuming that final identical pions are emitted at a system
configuration characterized by a freeze-out temperature $T_{\rm f}$,
we may generate the pion momenta according to Bose-Einstein
distribution and construct the single-event and mixed-event two-pion
correlation functions \cite{Zha05,Zha06} for NEXSPHERIO events.  Fig.
4 shows the two-pion correlation functions $C(q_{\rm side}, q_{\rm
out}, q_{\rm long})$ for the NEXSPHERIO events with different impact
parameters.  Here $q_{\rm side}$, $q_{\rm out}$, and $q_{\rm long}$
are the components of ``side", ``out", and ``long" of relative
momentum of pion pair \cite{Pra86,Ber88}, which are calculated in
the LCMS frame \cite{PHE02a}.  In each panel of Fig. 4, the dashed
lines give the correlation functions for a sample of different
single events and the solid line is for the mixed event obtained by
averaging over 40 single events. In the present calculations, the
freeze-out temperature is taken to be 150 MeV.  For each single
event the total number of generated pion pairs in the relative
momentum region ($q_{\rm side}, q_{\rm out}, q_{long} \leq 200$
MeV/c) is $N_{\pi\pi}=10^6$ and the numbers of the pion pairs in the
relative momentum regions $(q_i \leq 200\, {\rm MeV/c};\, q_j,\, q_k
\leq 30\, {\rm MeV/c})$ are about $2.7\%N_{\pi\pi}$, where $i$, $j$,
and $k$ denote ``side", ``out", and ``long".

\begin{figure}
\includegraphics[angle=0,scale=0.38]{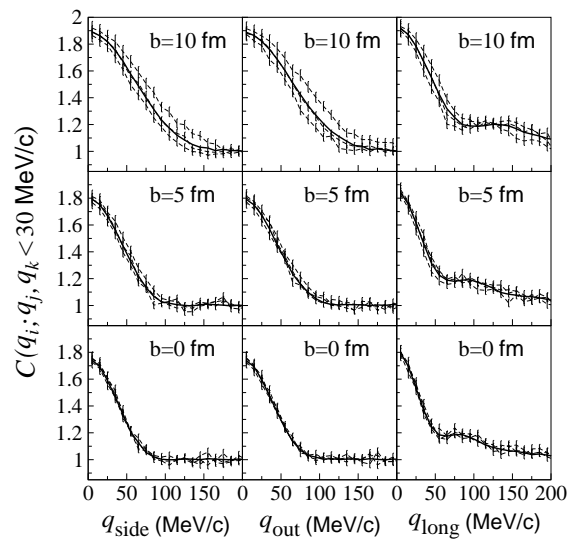}
\caption{\label{fig:fig4} Two-pion correlation functions for a sample of
different single events (dashed lines) and mixed events (solid lines) for
the NEXSPHERIO events with different impact parameters. }
\end{figure}

From Fig. 4 it can be seen that the correlation functions for single
events exhibit fluctuations relative to those for mixed events.
These fluctuations are larger for bigger impact parameter.  It is because
that the number of the lumps in system decreases with the impact parameter
and the fluctuations are larger for the source with smaller number of
lumps \cite{Won04,Zha05}.
It also can be seen that in the longitudinal direction the correlation
functions exhibit oscillations which can not be smoothed out by event
mixing.  This is because that there are two sub-sources moving forward
and backward the beam direction.  By applying an additional cut for the
initial rapidity of the ``smoothed particles", $\eta_0 >0$, we find that
the oscillations of the mixed-event correlation functions are smoothed
out as shown in Fig. 5(c).

\begin{figure}
\includegraphics[angle=0,scale=0.42]{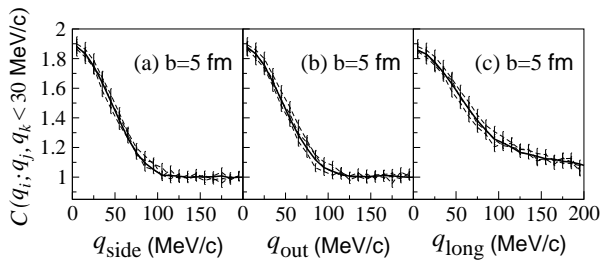}
\caption{\label{fig:fig5} Two-pion correlation functions for a sample
of different single events (dashed lines) and mixed events (solid lines)
with the initial rapidity cut $\eta_0 > 0$.}
\end{figure}

\section{Distribution of the fluctuation of single-event HBT
correlation function}

In last section we show that the two-pion correlation functions of
single events exhibit event-by-event fluctuations.  In realistic
experimental situations, the number of identical pion pairs in each
single event is limited.  Traditional HBT measurements are based on
mixed-event analysis because of statistics.  The event-by-event
fluctuations are smoothed out in the mixed-event analysis.  In order
to observe the event-by-event fluctuations, we next investigate the
distribution $dN/df$ of the fluctuations between the correlation
functions of single and mixed events, $|C_s(q_i) - C_m(q_i)|$, with
their error-inverses as weights \cite{Zha05},
\begin{eqnarray}
\label{RF}
f(q_i)  = \frac{|C_s(q_i) - C_m(q_i)|}{\Delta |C_s(q_i) - C_m(q_i)|} \,.
\end{eqnarray}
In the calculations, we take the width of the relative momentum $q_i$ bin
as 10 MeV and use the bins in the region $20 \le q_i \le 200$ MeV.
The up panels of Fig. 6 show the distributions of $f$ in the ``side",
``out", and ``long" directions, obtained from 40 simulated vents.
The impact parameter for these events is 5 fm ($b=5$ fm) and
the number of correlated pion pairs for each of these events is $10^7$
($N_{\pi\pi}=10^7$).  The thick lines are the results for the events with
fluctuating initial conditions (FIC).  Because in the last
analysis the fluctuations of the single-event correlation functions are
from the FIC, for comparison we also investigate the distribution $dN/df$
for the events with the smooth initial conditions (SIC) obtained by
averaging over 30 random NEXUS events \cite{Ham04,Soc04,Ham05} (although
it is not a realistic case).
It can be seen that the $f$ distributions for FIC are wider for the higher
freeze-out temperature $T_{\rm f}=150$ MeV than those for the lower $T_{\rm
f}=125$ MeV, and both for the freeze-out temperatures the distributions for
FIC are much wider than the corresponding results for SIC.
The down panels of Fig. 6 show the distributions of $f$ for 40 simulated
events with $b=5$ fm and $N_{\pi\pi}=5\times10^6$.  It can bee seen that
the distributions for FIC are also wider than those for SIC in this case.

\begin{figure}
\includegraphics[angle=0,scale=0.35]{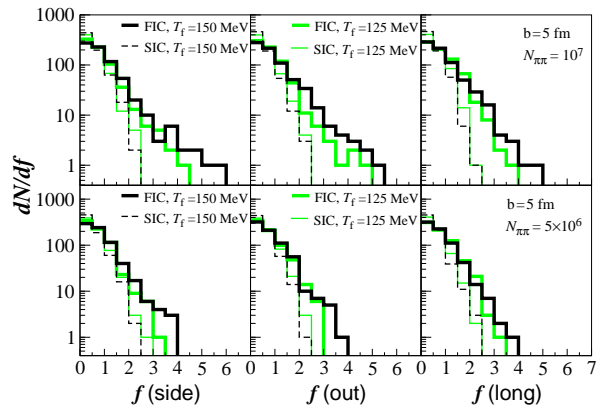}
\caption{\label{fig:fig6} The distributions $dN/f$ for 40 events
with FIC and SIC.  $b=5$ fm.}
\end{figure}

\begin{figure}
\includegraphics[angle=0,scale=0.48]{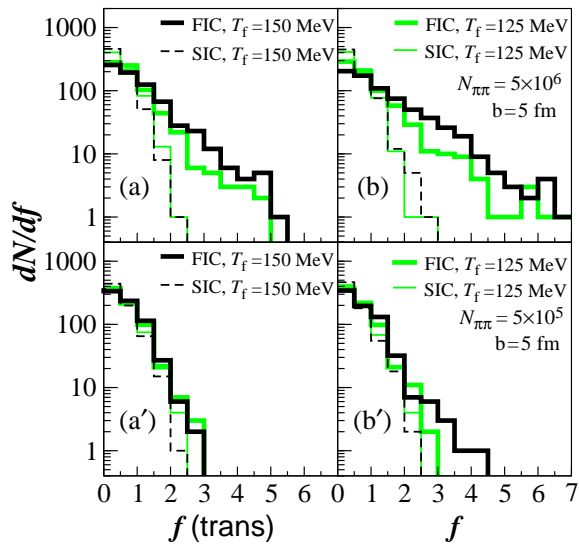}
\caption{\label{fig:fig7} The distributions $dN/df$ for 40 events,
where $f$ are calculated for the variables of $q_{\rm trans}$ and
$q$. $b=5$ fm.}
\end{figure}

In experiments the number of correlated pion pairs in one event,
$N_{\pi\pi}$, is limited.  It is related to the energy $\sqrt{s_{NN}}$ 
of the collisions.  For a finite $N_{\pi\pi}$, sometimes we have
to reduce variable numbers in data analyses although it will lose
some details.  In Fig. 7 we show the distributions of $f$ for the
variables of the transverse relative momentum $q_{\rm trans}$ and
the relative momentum $q$ for 40 events with $b=5$ fm.  One can see
that for $N_{\pi\pi}=5\times10^6$, the distributions for FIC are
much wider than those for SIC both for $q_{\rm trans}$ and $q$. Even
for $N_{\pi\pi}=5\times10^5$, the widths for FIC are visibly larger
than those for SIC for $q$.  In order to examine quantitatively, we
calculate the root-mean-square (RMS) of the $f$.  Figure 8 shows the
RMS $f_{\rm rms}$ as a function of $N_{\pi\pi}$ for the 40 events
with $b=9$ fm (the up panels) and $b=5$ fm (the down panels).  The 
freeze-out temperature is $T_{\rm f}=150$ MeV and the SIC are obtained 
by averaging over 100 NEXUS events.  It can been seen that the values 
of $f_{\rm rms}$ rapidly increase with $N_{\pi\pi}$ for FIC because 
the errors in Eq. (\ref{RF}) decrease with $N_{\pi\pi}$.  For FIC 
the results for $b=9$ fm are larger than those for $b=5$ fm 
correspondingly because the differences $|C_s(q_i) - C_m(q_i)|$ in 
Eq. (\ref{RF}) increase with $b$.  For SIC the values of $f_{\rm rms}$ 
are almost independent from $N_{\pi\pi}$ .  It is because that both the 
differences and their errors in Eq. (\ref{RF}) decrease with $N_{\pi\pi}$ 
in the SIC case.  

\begin{figure}
\includegraphics[angle=0,scale=0.65]{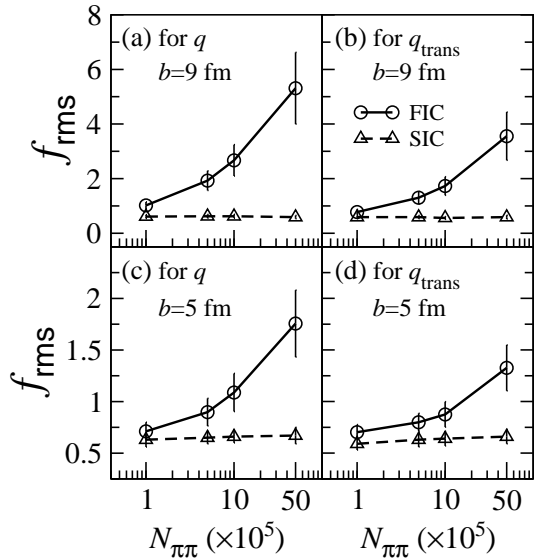}
\caption{\label{fig:fig8} The root-mean-square of $f$ for the
variables of $q$ and $q_{\rm trans}$ as a function of $N_{\pi\pi}$.
$T_{\rm f}=150$ MeV.}
\end{figure}

At RHIC energy the event multiplicity of identical pions, $M_{\pi}$,
is about several hundreds for central collisions.  The order of
$N_{\pi\pi}$ is about $10^5$ ($\sim M_{\pi}^2/2$).  However, at the higher
energy of the Large Hadron Collider (LHC) at CERN, $M_{\pi}$ will be 
about two thousands and the order of $N_{\pi\pi}$ will be $10^6$.  
Our results indicate that if the particle-emitting sources at LHC energy 
have the similar inhomogeneity as that at RHIC energy, the RMS of $f$ 
at LHC energy will be much larger than the corresponding result at RHIC 
energy.  The distributions of $f$ and their RMS are useful observables 
for the inhomogeneity of the sources.

\section{Conclusion}

In an event-by-event basis, the initial density distribution of
matter in nucleus-nucleus collisions have been known to be highly
fluctuating. The fluctuating initial conditions lead to
event-by-event inhomogeneous particle-emitting sources.  For these
sources the two-pion HBT correlation functions of single events
exhibit event-by-event fluctuations.  Because of data statistics
traditional HBT measurements are mixed-event analysis. In these
measurements the fluctuations of the correlation functions of single
events are smoothed out.  In order to observe the fluctuations we
examine the distributions of the error-inverse-weighted fluctuations 
between the HBT correlation functions of single and mixed events.  
We find that the distributions and the RMS of the weighted fluctuations 
are useful observables for the inhomogeneity of the sources.  If the 
particle-emitting sources produced in the heavy ion collisions at LHC 
energy are inhomogeneous, these RMS will be much larger than those at 
RHIC energy.

\begin{acknowledgments}
This research was supported by the National Natural Science Foundation of
China under Contracts No. 10575024 and No. 10775024.
\end{acknowledgments}

\end{document}